\documentclass{aa}

\usepackage{hyperref,amsmath}
\usepackage{booktabs}

\usepackage{graphicx}
\usepackage{txfonts}

\begin{document}

   \title{Fundamental physics with ESPRESSO: Constraining a simple parametrisation for varying $\alpha$}

\author{
Vitor da Fonseca \inst{1,2}\and
Tiago Barreiro \inst{1,3}\and
Nelson J. Nunes \inst{1,2}\and
Stefano Cristiani\inst{4,5}\and
Guido Cupani\inst{4,5}\and\\
Valentina D'Odorico\inst{4,5}\and
Ana C. O. Leite\inst{8,9}\and
Catarina M. J. Marques\inst{10,9}\and\\
Carlos J. A. P. Martins\inst{8,9}\and
Dinko Milakovi\'c\inst{4,5,11}\and
Paolo Molaro\inst{4,5}\and
Michael T. Murphy\inst{12,5}\and
Tobias M. Schmidt\inst{13,4}\and
Manuel Abreu\inst{1,2}\and
Vardan Adibekyan\inst{8,14}\and
Alexandre Cabral\inst{1,2}\and
Paolo Di Marcantonio\inst{4}\and\\
Jonay I. Gonz\'alez Hern\'andez\inst{6,7}\and
Enric Palle\inst{6,7}\and
Francesco A. Pepe\inst{13}\and
Rafael Rebolo\inst{6,7}\and
Nuno C. Santos\inst{8,14}\and\\
S\'ergio G. Sousa\inst{8,14}\and
Alessandro Sozzetti\inst{15}\and
Alejandro Su\'arez Mascare\~no\inst{6,7}\and
Maria-Rosa Zapatero Osorio\inst{16}
}

\institute{
Instituto de Astrof\'isica e Ci\^encias do Espa\c{c}o, Faculdade de Ci\^encias da Universidade de Lisboa, \\Campo Grande, PT1749-016 Lisboa, Portugal,\and
Departamento de Física da Faculdade de Ciências da Universidade de Lisboa, Edifício C8, 1749-016 Lisboa, Portugal\and
ECEO, Universidade Lus\'ofona de Humanidades e Tecnologias, Campo Grande, 376, 1749-024 Lisboa, Portugal\and
INAF -- Osservatorio Astronomico di Trieste, via G. B. Tiepolo 11, I-34143 Trieste, Italy\and
Institute for Fundamental Physics of the Universe, Via Beirut 2, I-34151 Miramare, Trieste, Italy\and
Instituto de Astrof\'{\i}sica de Canarias (IAC), Calle V\'{\i}a L\'actea s/n, E-38205 La Laguna, Tenerife, Spain\and
Departamento de Astrof\'{\i}sica, Universidad de La Laguna (ULL), E-38206 La Laguna, Tenerife, Spain\and
Instituto de Astrof\'isica e Ci\^encias do Espa\c co, CAUP, Universidade do Porto, Rua das Estrelas, 4150-762, Porto, Portugal\and
Centro de Astrof\'{\i}sica da Universidade do Porto, Rua das Estrelas, 4150-762 Porto, 
Portugal\and
Faculdade de Ci\^encias e Tecnologia, Universidade Nova de Lisboa, 2829-516 Caparica, Portugal\and
Istituto Nazionale di Fisica Nucleare, Sezione di Trieste, Via Bonomea 265, 34136 Trieste, Italy\and
Centre for Astrophysics and Supercomputing, Swinburne University of Technology, Hawthorn, Victoria 3122, Australia\and  
Observatoire Astronomique de l’Universit\'e de Gen\`eve, Chemin Pegasi 51, Sauverny 1290, Switzerland\and
Departamento de F\'isica e Astronomia, Faculdade de Ci\^encias, Universidade do Porto,\\ Rua Campo Alegre, 4169-007, Porto, Portugal\and
INAF -- Osservatorio Astrofisico di Torino, via Osservatorio 20, 10025 Pino Torinese, Italy\and
Centro de Astrobiolog\'{\i}a (CSIC-INTA), Crta. Ajalvir km 4, E-28850 Torrej\'on de Ardoz, Madrid, Spain
}
   \date{}
   
    \abstract{The spectrograph ESPRESSO recently obtained a limit on the variation of the fine-structure constant, $\alpha$, through measurements along the line of sight of a bright quasar with a precision of  $1.36$ ppm at $1\sigma$ level. This imposes new constraints on cosmological models with a varying $\alpha$. We assume such a model where the electromagnetic sector is coupled to a scalar field dark energy responsible for the current acceleration of the Universe. We parametrise the variation of $\alpha$ with two extra parameters, one defining the cosmological evolution of the quintessence component and the other fixing the coupling with the electromagnetic field.}
{The objective of this work is to constrain these parameters with both astrophysical and local probes. We also carried out a comparative analysis of how each data probe may constrain our parametrisation.}
{We performed a Bayesian analysis by comparing the predictions of the model with observations. The astrophysical datasets are composed of quasar spectra measurements, including the latest ESPRESSO data point, as well as Planck observations of the cosmic microwave background. We combined these with local results from atomic clocks and the MICROSCOPE experiment.}
{The constraints placed on the quintessence parameter are consistent with a null variation of the field, and are therefore compatible with a $\Lambda$CDM cosmology. The constraints on the coupling to the electromagnetic sector are dominated by the E\"otv\"os parameter local bound.}
{More precise measurements with ESPRESSO will be extremely important to study the cosmological evolution of $\alpha$ as it probes an interval of redshift not accessible to other types of observations. However, for this particular model, current available data favour a null variation of $\alpha$ resulting mostly from the strong MICROSCOPE limits.}
 

   \maketitle

\section{\label{sec:intro}Introduction}

Like the other fundamental constants in physics, the value of the fine-structure constant, $\alpha\equiv e^2/\hbar c$, which defines the strength of the electromagnetic interaction, is only determined through experimental measurements in the absence of theoretical predictions. While $\alpha$ has been measured in the lab with extremely high precision, whether or not it varies in space or time is still unknown. The detection of any variation would open prospects beyond the standard model of particle physics \citep{uzan2011}.

Locally, strong constraints are obtained on the drift rate of the fine-structure constant by  comparing pairs of atomic clocks. The experiments are based on the fact that atomic transitions have a frequency that manifests a different dependency on the value of $\alpha$. Usually, various transitions are measured in different atoms over a given time span. However, \citet{Lange2021} obtained the most stringent limit on a temporal drift of the fine-structure constant by comparing two different transition frequencies provided by the same ytterbium ion over a period of about 4 years. The authors improved the limits for the temporal variations of $\alpha$ to $1.0(1.1)\times10^{-18}/$yr.

On cosmological timescales, the observation of distant quasi-stellar object (QSO) spectra has proven beneficial for testing the stability of the fine-structure constant for more than 20 years. The measurements of absorption lines provide a means to directly determine $\alpha$ in intervening extragalactic gas clouds that correspond to different look-back times. Data from Keck/HIRES suggested that $\alpha$ was smaller in the past at redshift $0.5<z<3$ \citep{2001webb,murphy2003} whereas an initial analysis of observations from the southern hemisphere with VLT/UVES did not find any indication of such fluctuation \citep{chand2004}. However, a re-analysis of the UVES spectra, which corrected the absorption-profile-fitting procedure, produced results consistent with the Keck ones \citep{Murphy2007,Murphy2008}; though see also \citet{https://doi.org/10.48550/arxiv.0711.1742}. Subsequent analysis of archival data from both HIRES and UVES spectrographs successfully reconciled the results from the two hemispheres by invoking possible spatial variation of the fine-structure constant across the sky \citep{webb2011, king2012}. Nevertheless, the Keck and VLT spectra suffer from wavelength calibration errors that undermine the conclusions \citep{2014whitemore}.  In contrast to the archival data, which was not collected for the specific purpose of measuring $\alpha$, recent dedicated measurements made with UVES and HIRES that mitigate wavelength calibration errors are compatible with the null result \citep{martins2017,2019PDU....25..301M,2022PhLB..82737002M}.

It is in this context of technical challenges and stimulating controversies  that the spectrograph ESPRESSO \citep{espresso2021} for the VLT was designed with the specific aim of testing the fine-structure constant stability, unlike the older high-resolution spectrographs. A highly stabilised environment, the fibre feed, and the possibility of calibration with a femtosecond-pulsed laser frequency comb (LFC) make it possible to accurately measure the absorption features with an exquisite control of systematic errors,
effectively removing wavelength calibration from the error budget. The first ESPRESSO measurements provide a new constraint on relative variations in $\alpha$, namely $\Delta\alpha/\alpha=1.31\pm 1.36$ ppm at redshift $z=1.15$, which is consistent with no cosmological variation \citep{2022espresso}. Nevertheless, this additional result is useful for further testing of cosmological models that predict varying fundamental constants.

This paper considers one particular class of these models, where the dark-energy component is in the form of quintessence assumed to be non-minimally coupled to the electromagnetic sector \citep[see][]{martins2017}. Indeed, almost massless scalar fields necessarily couple to matter, provided that the coupling is not suppressed by a symmetry \citep{carroll1998}. The theory supposes a varying fine-structure constant by coupling the scalar field, $\phi$, to the electromagnetic Faraday tensor as $B_F(\phi)F_{\mu\nu}F^{\,\mu\nu}$, where $B_F$ is the so-called gauge kinetic function \citep{PhysRevD.25.1527,copeland2004}. Accordingly, the fine-structure constant evolves as $\alpha\propto B_F^{-1}(\phi)$, and therefore depends on the value of the scalar field which is responsible for the late time acceleration of the Universe expansion \citep{acel1,acel2}. Our aim is to constrain our model with the help of various observational datasets. To this end, we assume a linear dependency of the gauge kinetic function on the scalar field, such that $B_F(\phi)=1-\zeta\kappa(\phi-\phi_0)$, where $\zeta$ is the coupling constant. Moreover, we use a simple parametrisation for dark energy where the scalar field linearly scales with the number of e-folds \citep{Nunes:2003ff}, $\phi^\prime=\lambda$, for some constant $\lambda$ (the prime denotes the derivative with respect to the number of e-folds).  The two slopes $\zeta$ and $\lambda$ of the linear dependencies are free parameters of the model.

We made a joint constraint of the two parameters with data from both QSO spectral lines and atomic clocks, although they can also be separately constrained. As the light scalar field couples to matter, it entails the violation of the universality of free fall \citep{Damour2010}. This kind of model is therefore extremely constrained from the bound on the weak equivalence principle (WEP). The level of WEP violation can be quantified with the E\"otv\"os parameter $\eta$ which relates to $\zeta$ in a model-dependent way \citep{carroll1998,Dvali2002,2002PThPh.107..631C}. We made use of the bound on $\eta$ obtained with the MICROSCOPE space mission (MICROSatellite pour l’Observation du Principe d’Equivalence) in \citet{Touboul2019} to individually constrain $\zeta$.  The MICROSCOPE experiment is designed to confirm the equivalence between inertial mass and gravitational mass by observing the free-fall motion of two masses ---made of two different materials--- sourced by the Earth's gravitational field. For $\lambda$, we used a distance prior based on existing cosmic microwave background (CMB) constraints for dark energy. This prior imposes stringent cosmological limits on the evolution of the scalar field from a high-redshift probe.

\section{\label{sec:Theory}Description of a varying $\alpha$ cosmological model}

We choose to extend the spatially flat $\Lambda$CDM model by generalising the cosmological constant $\Lambda$ with a varying canonical and homogeneous scalar field accounting for dark energy; the scalar field is minimally coupled to gravity but non-minimally coupled to the electromagnetic field. Assuming the same matter Lagrangian, $\mathcal{L}_M$, as in the standard model composed of radiation, baryons, and cold dark matter, the overall action that characterises this theory is given by
\begin{equation}
\label{eq:action}
S=-\frac{1}{2\kappa^2}\int R\sqrt{-g}\,d^{\,4}x+\int\left(\mathcal{L}_M+\mathcal{L}_\phi+\mathcal{L}_{\phi F}\right)\sqrt{-g}\,d^{\,4}x\,,
\end{equation}
where $R$ is the Ricci scalar, $g$ is the determinant of the metric $g_{\mu\nu}$ , and $\kappa^2\equiv8\pi G/c^4$ is the gravitational coupling constant. The Lagrangian density for the quintessence component is given by
\begin{equation}
\label{eq:lagrangian_quintessence}
\mathcal{L}_\phi=\frac{1}{2}\partial^{\,\mu}\phi\partial_\mu\phi-V\left(\phi\right)\,,
\end{equation}
while its interaction with the electromagnetic field is represented by the following Lagrangian density:
\begin{equation}
\label{eq:lagrangian_electromagnetic}
\mathcal{L}_{\phi F}=-\frac{1}{4}B_F\left(\phi\right)F_{\mu\nu}F^{\,\mu\nu}\,,
\end{equation}
where $V(\phi)$ is the potential of the scalar field, $F_{\mu\nu}$ is the strength of the electromagnetic field, and $B_F(\phi)$ is the gauge kinetic function which defines the coupling between dark energy and the electromagnetic sector. In this model, the coupling imposes the variation of the fine-structure constant whose cosmological evolution is given by
\begin{equation}
\label{eq:variation_alpha}
\alpha\left(\phi\right)=\alpha_0 B_F^{-1}\left(\phi\right)\,,
\end{equation}
where $\alpha_0\approx1/137$ is the present-day value. As the possible variations in $\alpha$ are necessarily small, one can harmlessly assume that the Taylor expansion of $B_F(\phi)$ at first order is valid on cosmological timescales,
\begin{equation}
\label{eq:BF_expansion}
B_F\left(\phi\right)=1-\zeta\kappa\left(\phi-\phi_0\right)\,,
\end{equation}
where $\phi_0$ is the field present value and $\zeta$ is a dimensionless constant that parametrises the linear dependency of the gauge kinetic function on the scalar field. It follows from Eq.~(\ref{eq:variation_alpha}) that $\zeta$ is also the slope of the linear dependency of the effective fine-structure constant on the scalar field,
\begin{equation}
\label{eq:alpha_phi}
\frac{\Delta\alpha}{\alpha}\equiv\frac{\alpha-\alpha_0}{\alpha_0}=\zeta\kappa\left(\phi-\phi_0\right)\,.
\end{equation}

For testing this model, we use the parametrisation for dark energy of \citet{Nunes:2003ff} where quintessence scales with the number of e-folds, $N\equiv\ln a$,
\begin{equation}
\label{eq:lambda_param}
\kappa\left(\phi-\phi_0\right)=\lambda N\,,
\end{equation}
for a given constant $\lambda$.  Logarithmic dependencies on the scale factor exist in dilaton-type models inspired by string theories in which $\lambda$ has different values in the acceleration, matter, and radiation eras \citep{Damour2002}. The scalar field potential, $V(\phi)$, derived in \citet{Nunes:2003ff}, takes the form of a double exponential which is able to bring the Universe towards its late time acceleration almost irrespectively of the initial conditions \citep{Barreiro:1999zs}:
\begin{equation}
\label{eq:potential}
V\left(\phi\right)=A\,e^{-\frac{3}{\lambda}\kappa\phi}+B\,e^{-\lambda\kappa\phi}\,,
\end{equation}
where the mass scales $A$ and $B$, normalised to $\phi_0=0$, are completely fixed by the cosmological parameters
\begin{align}
A &=\frac{\lambda^2}{3-\lambda^2}\frac{3H_0^2}{2\kappa^2}\,\Omega_{m},\\
B &=\frac{\lambda^2-6}{3-\lambda^2}\,\frac{3H_0^2}{2\kappa^2}\left(\frac{\lambda^2}{3}-1+\Omega_{m}\right)\,,
\label{eq:mass_scales}
\end{align}
where $H_0$ and $\Omega_m$ are respectively the current Hubble expansion rate and today's abundance of matter. 

We were motivated in the choice of this parametrisation by the fact that it appropriately helps to constrain the dark energy equation of state, $w$. It also captures a wide range of evolution at low redshift with only one extra degree of freedom \citep{DAFONSECA2022100940}. In this respect, it usefully complements the familiar CPL parametrisation \citep{param_a1,param_a2}. Our additional parameter $\lambda$ sets today's dark energy equation of state:
\begin{equation}
\label{eq:w0}
w_0=-1+\frac{\lambda^2}{3\left(1-\Omega_m\right)}\,.
\end{equation}
This dark energy parametrisation possesses a $\Lambda$CDM limit when $w_0\rightarrow-1$ for $\lambda\rightarrow0$ , allowing a direct test of departures from the concordance model.  At higher redshift, during matter domination, $w\rightarrow0$, making it by construction a well-behaved model \citep[see][]{Nunes:2004eog}. According to the parametrisation, the cosmological variation of $\alpha$ in Eq.~(\ref{eq:alpha_phi}) becomes
\begin{equation}
\frac{\Delta\alpha}{\alpha}=\zeta\lambda\ln a=-\zeta\lambda\ln\left(1+z\right)\,.
\label{eq:alpha_param}
\end{equation}

Thus we obtain a model in which the relative variation in $\alpha$ scales with the number of e-folds, the amplitude of the linear dependency being the product of the two additional parameters, $\zeta\lambda$. The value of the fine-structure constant therefore logarithmically depends on redshift. This model only adds two extra degrees of freedom to the standard model, limiting the degeneracies between cosmological parameters. As the two parameters are perfectly degenerated, we constrain the model in the following section by adding priors and external datasets to the observations that specifically constrain $\Delta\alpha/\alpha$.

\section{\label{sec:Data}Selection of observational data to constrain the model}

Two observables sensitive to the product $\zeta\lambda$ are available to constrain the two parameters jointly. Firstly, the predictions of Eq.~(\ref{eq:alpha_param}) can be compared to astrophysical measurements in extragalactic clouds based on QSO absorption spectra, and secondly, we used the current drift rate of the fine-structure constant constrained by atomic clocks.

\begin{table*}[t]
\caption{QSO dataset. The first column is the name of the source, $z_\textrm{abs}$ is the absorber redshift of the measurement {along the line of sight}, $\Delta\alpha/\alpha$ is the respective measured variation of the fine-structure constant in parts per million (ppm), with the corresponding quadrature sum of statistical uncertainty and systematic error ($1\sigma$). The two last columns give respectively the name of the spectrograph used and the reference of the measurement.}
\label{tab:QSO}
\centering
\begin{tabular} {c c c c c}
\hline\hline
Quasar &$z_\textrm{abs}$ & $\Delta\alpha/\alpha$ & Spectrograph & Reference\\
& & \multicolumn{1}{c}{(ppm)} \\
\hline
J0120{\ensuremath{+}}2133 & $0.729$ & $0.73\pm6.42$ & HDS & 1 \\
J0026{\ensuremath{-}}2857& $1.023$ & $3.54\pm8.87$ & UVES & 2 \\
J0058{\ensuremath{+}}0041 & $1.072$ & $-1.35\pm7.16$ & HIRES & 2 \\
3 quasars & $1.080$ & $4.30\pm3.40$ & HIRES & 3 \\
HS1549{\ensuremath{-}}1919 & $1.143$ & $-7.49\pm5.53$ & UVES/HIRES/HDS & 4 \\
HE0515{\ensuremath{-}}4414 & $1.151$ & $1.31\pm1.36$ & ESPRESSO & 5 \\
HE0515{\ensuremath{-}}4414 & $1.151$ & $-1.42\pm0.85$ & UVES & 6 \\
HE0515{\ensuremath{-}}4414 & $1.151$ & $-0.27\pm2.41$ & HARPS & 7 \\
J1237{\ensuremath{+}}0106 & $1.305$ & $-4.54\pm8.67$ & HIRES & 2 \\
J0120{\ensuremath{+}}2133 & $1.325$ & $2.60\pm4.19$ & HDS & 1 \\
HS1549{\ensuremath{+}}1919 & $1.342$ & $-0.70\pm6.61$ & UVES/HIRES/HDS& 4 \\
J0841{\ensuremath{+}}0312 & $1.342$ & $3.05\pm3.93$ & HIRES & 2 \\
J0841{\ensuremath{+}}0312 & $1.342$ & $5.67\pm4.71$ & UVES & 2 \\
 J0120{\ensuremath{+}}213 & $1.343$ & $8.36\pm12.16$ & HDS & 1 \\
J0108{\ensuremath{-}}0037& $1.371$ & $-8.45\pm7.34$ & UVES & 2 \\
HE0001{\ensuremath{-}}2340& $1.580$ & $-1.50\pm2.60$ & UVES & 8  \\
J1029{\ensuremath{+}}1039& $1.622$ & $-1.70\pm10.11$ & HIRES & 2 \\
HE1104{\ensuremath{-}}1805& $1.661$ & $-4.70\pm5.30$& HIRES & 3 \\
HE2217{\ensuremath{-}}2818 & $1.692$ & $1.30\pm2.60$ & UVES & 9 \\
HS1946{\ensuremath{+}}7658 & $1.738$ & $-7.90\pm6.20$ & HIRES & 3 \\
HS1549{\ensuremath{+}}1919 & $1.802$ & $-6.42\pm7.25$ & UVES/HIRES/HDS & 4 \\
Q1103{\ensuremath{-}}2645& $1.839$ & $3.30\pm2.90$ & UVES & 10 \\
Q2206{\ensuremath{-}}1958 & $1.921$ & $-4.65\pm6.41$ & UVES & 2 \\
Q1755{\ensuremath{+}}57 & $1.971$ & $4.72\pm4.71$ & HIRES & 2 \\
PHL957 & $2.309$ & $-0.65\pm6.84$ & HIRES & 2 \\
PHL957 & $2.309$ & $-0.20\pm12.93$ & UVES & 2 \\
J0035{\ensuremath{-}}0918 & $2.340$ & $-12.0\pm11.0$ & ESPRESSO & 11 \\
\hline
\end{tabular}
\tablebib{
(1)~\citet{MurphyCooksey2017}; (2) \citet{Murphy2016}; (3) \citet{SongailaCowie2014}; (4) \citet{Evans2014};
(5) \citet{2022espresso}; (6) \citet{Kotus2016}; (7) \citet{Milakovic2020}; (8) \citet{Agafonova2011};
(9) \citet{molaro2013}; (10) \citet{BainbridgeWebb2017}; (11) \citet{Welsh2020}.
}
\end{table*}
\begin{figure}[t]
 \resizebox{\hsize}{!}{\includegraphics{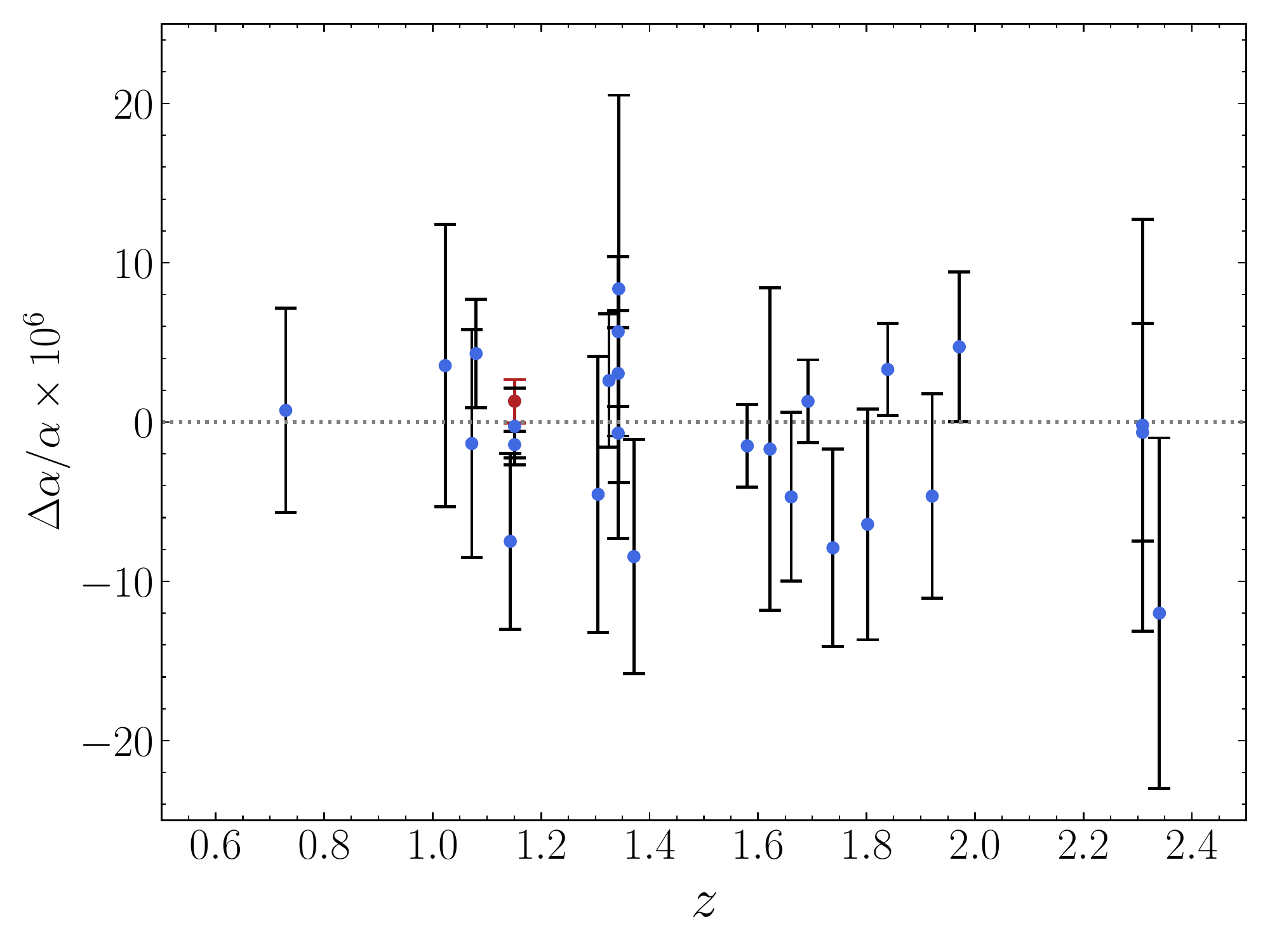}}
  \caption{Measurements of the QSO dataset.The most recent ESPRESSO data point is shown in red \citep{2022espresso}.}
\label{fig:QSO}
\end{figure}

As for the astrophysical measurements, our dataset referred to as `QSO' is analogous to that of \citet{Martinelli2021} to ensure comparability with previous works similar to the present one, where it is common practice to test cosmological models with a set of dedicated and well-calibrated absorption line measurements of $\Delta\alpha/\alpha$. We refer the reader to Sect.~3 of \citet{martins2017} for a thorough review. The dataset consists of 27 measurements in systems that are located at different redshifts along several quasar lines of sight (see Table~\ref{tab:QSO}). These allow us to place constraints on the value of the fine-structure constant up to redshift $z=2.34$ as illustrated by Fig.~\ref{fig:QSO}. In order to ensure some degree of homogeneity of the sample, the dataset is limited to those measurements that were well calibrated and specifically taken to test the stability of $\alpha$, including the most recent ESPRESSO result \citep{2022espresso}. The 27 measurements were obtained under dedicated conditions whereby the data acquisition procedures ensured tailored wavelength calibration and control of systematic errors.

In particular, the result from redshift ${z=1.692}$ was obtained with the HE2217{\ensuremath{-}}2818 spectrum \citep{molaro2013} under the UVES Large Program for testing fundamental physics. The program was designed to test the possible cosmological variations in $\alpha$ that were previously detected through the analysis of the archival statistical sample of quasar spectra composed of 293 Keck/HIRES and VLT/UVES measurements \citep{webb2011}. It involved several observational groups to improve the calibration and optimise the data reduction and analysis pipeline. Under the same program, the separate observation of the equatorial quasar HS1549{\ensuremath{+}}1919 by VLT, Keck, and Subaru \citep{Evans2014} led to the removal of the long-range distortions in the wavelength scales of the quasar spectra affecting the UVES, HIRES, and HARPS spectrographs   from the systematic errors \citep{2014whitemore}. This improvement was achieved by comparing the spectra obtained with the three telescopes and using `supercalibration' techniques. The corresponding data points in Table~\ref{tab:QSO} are the average of the three measurements from each absorber system, ${z=1.143}$, ${z=1.342,}$ and ${z=1.802,}$ respectively.

Likewise, \citet{MurphyCooksey2017} followed a supercalibration procedure to correct the long-range distortions in two Subaru quasar spectra. A blinded analysis provided the following three measurements from absorbers where $\Delta\alpha/\alpha$ is relatively well constrained: $z = 0.729$, $1.325,$ and $1.343$ towards J0120{\ensuremath{+}}2133. Furthermore, instead of using the common iron or magnesium quasar absorptions lines, precise limits on the variability of $\alpha$ were obtained by \cite{Murphy2016} with zinc and chromium lines which are the most sensitive to its relative variation, and also more resistant to long-range distortions of the wavelength calibration. These are measurements taken from the following absorber positions: ${z=1.023}$ towards J0226{\ensuremath{-}}2857, ${z=1.072}$ towards J0058{\ensuremath{+}}0041, ${z=1.305}$ towards J1237{\ensuremath{+}}0106, ${z=1.371}$ towards J0108{\ensuremath{-}}0037,  ${z=1.622}$ towards J1029{\ensuremath{+}}1039, ${z=1.921}$ towards Q2206{\ensuremath{-}}1958, and ${z=1.971}$ towards Q1755{\ensuremath{+}}57. The QSO dataset also includes two measurements from ${z=1.342}$ towards J0841{\ensuremath{+}}0312 and two measurements from ${z=2.309}$ towards PHL957 because they were independently taken using different telescopes and accepted into the main results of \cite{Murphy2016}.

By correcting the UVES spectra of HE0515{\ensuremath{-}}4414 with well-calibrated spectra from the High Accuracy Radial velocity Planet Searcher (HARPS), \cite{Kotus2016} obtained a measurement of high sensitivity from an intervening absorption system at $z=1.151$. Additionally, we make use of the objective measurement obtained in \citet{BainbridgeWebb2017} by applying artificial intelligence to automatise the analysis pipeline of the absorption spectra from the ${z=1.839}$ absorber towards Q1103{\ensuremath{-}}2645. We also include in the sample the measurement at ${z=1.080}$ of \cite{SongailaCowie2014} where careful attention was paid to the wavelength calibration and control of systematic errors to aim for precision in measuring $\Delta\alpha/\alpha$. The data point is the result of the weighted average from eight absorbers towards HE1104{\ensuremath{-}}1805A, HS1700{\ensuremath{+}}6416, and HS1946{\ensuremath{+}}7658 (the values for the  individual systems were not given). The dataset contains an older measurement from $z=1.580$ in the line of sight of quasar HE0001{\ensuremath{-}}2340 obtained by \cite{Agafonova2011} who studied the extent  to which the Mg II lines used as anchor were affected by isotopic shifts and tested the local accuracy of the wavelength scale calibration. It should be noted that \citet{Noterdaeme_2021} concluded that the strong enhancement of heavy Mg isotopes was probably due to distortion in the UVES wavelength scale.

More recent measurements are also taken into account: one from science verification observations obtained with ESPRESSO and calibrated with LFC from ${z=2.340}$ towards J0035{\ensuremath{-}}0918 \citep{Welsh2020}, and another from new observations of the $z=1.151$ absorption system along the HE0515{\ensuremath{-}}4414 sightline aided by LFC for the HARPS wavelength calibration \citep{Milakovic2020}. Finally, we take account of the first precise ESPRESSO measurement with LFC calibration \citep{2022espresso}, as the main motivation of the present paper, from $z=1.151$ towards HE0515{\ensuremath{-}}4414. As in the case of the absorbers towards J0841+0312 and PHL957, for which two different data points are used, we keep three measurements towards HE0515{\ensuremath{-}}4414 because they can be considered independent in the statistical analysis. The three measurements were obtained using different spectrographs (HARPS, UVES and ESPRESSO), yet with appropriate wavelength calibration and control of systematic errors.

We define the likelihood of the QSO dataset, namely the probability of getting the $\Delta\alpha/\alpha$ measurements given our theory, as follows:
\begin{equation}
\label{eq:QSO_likelihood}
\ln\mathcal{L}_\textrm{\tiny{QSO}}=-\frac{1}{2}\sum\limits_{i}\frac{1}{\sigma_i^2}\left[\left.\frac{\Delta\alpha}{\alpha}\right|_\textrm{\tiny{th}}\left(z_i\right)-\left.\frac{\Delta\alpha}{\alpha}\right|_\textrm{\tiny{obs}}\left(z_i\right)\right]^2\,,
\end{equation}
where $i$ denotes the $i^{\textrm{\tiny{th}}}$ observation in Table~\ref{tab:QSO}, $\sigma_i$ and $z_i$ are respectively the corresponding standard deviation and absorber redshift. The subscripts `th' and `obs' stand for the theoretical prediction of Eq.~(\ref{eq:alpha_param}) and the $i^{\textrm{\tiny{th}}}$ measurement, respectively.

Regarding the current $\alpha$ drift rate, $(\dot{\alpha}/\alpha)_0$, according to our parametrisation it simplifies to
\begin{equation}
\left(\frac{\dot{\alpha}}{\alpha}\right)_0=\zeta\lambda H_0 = \zeta\lambda h\,\left(1.02\, \times 10^{-10} \right)\,{\rm yr^{-1}}\,,
\label{eq:drift_rate}
\end{equation}
where the dot denotes derivation with respect to cosmic time and $h$ is the Hubble parameter. We note that there is a degeneracy with the current expansion rate, $H_0\equiv100h$ km/s/Mpc. The drift rate is locally constrained in lab experiments based on the comparison of atomic clocks, the strongest bound being in \citet{Lange2021},
\begin{equation}
\left(\frac{\dot{\alpha}}{\alpha}\right)_0=\left(1.0\pm1.1\right)\times 10^{-18}\,\,\textrm{yr}^{-1}\,.
\label{eq:drift_rate_bound}
\end{equation}
We refer to this constraint as `atomic clocks'. The corresponding likelihood is simply
\begin{equation}
\label{eq:clocks_likelihood}
\ln\mathcal{L}_\textrm{\tiny{clocks}}=-\frac{1}{2}\frac{\left[\left.\left(\frac{\dot{\alpha}}{\alpha}\right)_0\right|_\textrm{\tiny{th}}-\left.\left(\frac{\dot{\alpha}}{\alpha}\right)_0\right|_\textrm{\tiny{obs}}\right]^2}{\sigma^2}\,,
\end{equation}
where again the subscripts `th' and `obs' stand for the theoretical prediction of Eq.~(\ref{eq:drift_rate}) and the bound in Eq.~(\ref{eq:drift_rate_bound}) respectively, while $\sigma$ is the error in the measurement.

Moreover, it is possible to break the degeneracies between $\zeta$ and $\lambda$ by imposing individual constraints on them. To this end, we use the E\"otv\"os parameter, $\eta$, as observable to separately constrain the electromagnetic coupling $\zeta$ by considering the following relation \citep{carroll1998,Dvali2002}:
\begin{equation}
\eta\approx 10^{-3}\zeta^2\,.
\label{eq:eta}
\end{equation}
The parameter $\eta$ quantifies the level of violation of the weak equivalence principle induced by the variation of $\alpha$ \citep{Damour2010}. We make use of the local bound on $\eta$ in \citet{Touboul2019} placed by the space-based MICROSCOPE experiment, as it is the most stringent:
 \begin{equation}
\eta=\left(-0.1\pm1.3\right)\times 10^{-14}\,.
\label{eq:eta_bound}
\end{equation}
We call this bound `MICROSCOPE', for which we define the likelihood such that
\begin{equation}
\label{eq:microscope_likelihood}
\ln\mathcal{L}_\textrm{\tiny{MICRO}}=-\frac{1}{2}\frac{\left(\eta_\textrm{\tiny{micro}}-\eta\right)^2}{\sigma^2}\,,
\end{equation}
where $\eta_\textrm{\tiny{micro}}$ and $\sigma$ are given by Eq.~(\ref{eq:eta_bound}). Moreover, this latter is a Gaussian prior put on $\zeta$ in the statistical analysis.

As for $\lambda$, to ease the calculation, we constructed a multivariate Gaussian prior that approximatively summarises the geometrical constraints inferred in \citet{DAFONSECA2022100940}, where the present scalar field parametrisation was tested with Planck data, regardless of the electromagnetism coupling. We therefore apply it on the vector of free parameters, $\nu=(\lambda,\omega_b,\omega_c,H_0)$, without $\zeta$, where $\omega_b\equiv\Omega_bh^2$ is today's baryon density and $\omega_c\equiv\Omega_ch^2$ is today's density of cold dark matter\footnote{\citet{DAFONSECA2022100940} additionally contemplate a coupling within the dark sector parametrised with a constant $\beta$, which we consider to be vanishing  here.}. Our Gaussian prior, which is analogous to \citet{Betoule2014}, has the following likelihood form
\begin{equation}
\label{eq:gaussian_prior}
\ln\mathcal{L}_\textrm{\tiny{cmb}}=-\frac{1}{2}\left(\nu-\nu_\textrm{\tiny{cmb}}\right)^\top C_\textrm{\tiny{cmb}}^{-1}\left(\nu-\nu_\textrm{\tiny{cmb}}\right)\,,
\end{equation}
where $\nu_\textrm{\tiny{cmb}}=(\lambda,\omega_b,\omega_c,H_0)_\textrm{\tiny{cmb}}=(0.,0.02267,0.1210,66.8)$, and $C_\textrm{\tiny{cmb}}$ is the best-fit covariance matrix marginalised over all the fixed parameters, taken from the previous study carried out in \citet{DAFONSECA2022100940}:
\begin{equation}
\label{eq:planck_prior}
{\small C_\textrm{\tiny{cmb}}=
\begin{pmatrix}
3.0\times10^{-3} & -9.8\times10^{-9} & -2.2\times 10^{-6} & 1.8\times 10^{-3} \\
-9.8\times 10^{-9} & 3.6\times 10^{-8} & -1.8\times 10^{-8} & -2.1\times 10^{-5} \\
-2.2\times 10^{-6} & -1.8\times 10^{-8} & 8.9\times10^{-6} & -5.5\times 10^{-3} \\
1.8\times 10^{-3} & -2.1\times 10^{-5} & -5.5\times 10^{-3} & 3.6
\end{pmatrix}}\,.
\end{equation}
We refer to this likelihood as `Planck prior'.

\begin{figure}[t]
\resizebox{\hsize}{!}{\includegraphics{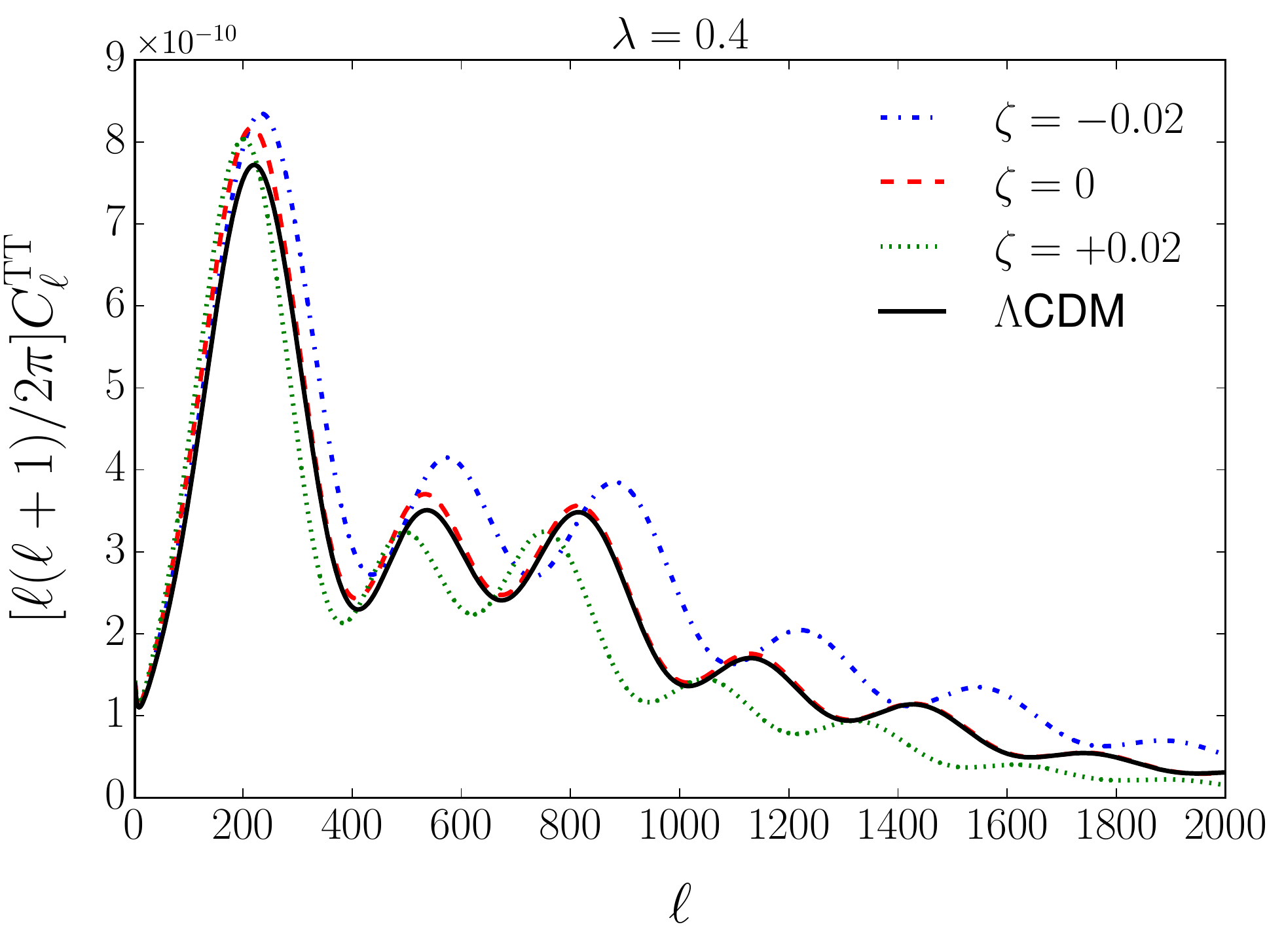}}
  \caption{CMB temperature power spectra for the dark energy parameter $\lambda=0.4$ and different values of the electromagnetic coupling $\zeta$.}
\label{fig:cmb}
\end{figure}

It is worthwhile noting that the Planck prior is built from a dark energy model that does not foresee any coupling with the electromagnetic sector. However, it was demonstrated that the CMB angular power spectrum is most sensitive to the value of the fine-structure constant at recombination time because varying $\alpha$ changes the ionisation history in the early Universe \citep{Kaplinghat1999}. The effects on the anisotropies are degenerate with other cosmological parameters whereas we limit the CMB constraints to a distance prior that does not take into account the variation of $\alpha$. It is therefore important to assess whether our approximation is accurate enough. To do so, we adapted the Einstein-Boltzmann code CLASS \citep{class2} to accommodate the potential of our specific scalar field parametrisation in Eq~.~\eqref{eq:potential}, and to include the evolution of the variation of the fine-structure constant obtained in Eq.~\eqref{eq:alpha_param}. We used this code to compute the angular power spectra of the CMB.  Figure~\ref{fig:cmb} shows the competing effects of our two parameters $\lambda$ and $\zeta$ on the position and amplitude of the acoustic peaks, when the other cosmological parameters are fixed to Planck 2018 values \citep{Planck2020}. On the one hand, as the presence of the scalar field speeds up the expansion of the Universe background, the comoving angular distance to the last scattering surface decreases with respect to $\Lambda$CDM. The increase in the Hubble rate slightly shifts the anisotropies towards larger scales. Moreover, the amplitude of the peaks is larger in comparison with the concordance model. The non-negligible contribution of dark energy in the early Universe reduces the fractional energy density of matter at the time of decoupling. On the other hand, as recapitulated in \citet{Hart2017}, larger $\alpha$ shifts the peaks towards smaller scales because of the earlier recombination, and enhances their amplitude via the suppression of photon diffusion damping. The opposite effects hold for decreasing $\alpha,$ whereas $\lambda$ influences the power spectrum irrespective of its sign.

In light of the examination of the degeneracies, we find it necessary in the following section to compare statistical results between the Planck prior that is insensitive to $\zeta$ and a more comprehensive Planck likelihood, referred to as `full Planck' hereafter,  which is sensitive to $\alpha$ variation. We used the legacy likelihood Planck TT, TE, EE described in \cite{Plancklikelihhod2020}, where TT, TE, and EE stand for the angular power spectra of the temperature auto-correlation, the temperature and polarisation cross-correlation, and the polarisation auto-correlation,  respectively. This derives from the data collected by the Planck satellite on the anisotropies of the CMB in the form of  Stokes intensity and linear polarisation maps complemented by intensity maps. The information is split into temperature T and the E mode polarisation component to construct the likelihood of a given theoretical model power spectrum given the observed one, taking account of correlations between the TT, TE, and EE spectra for the higher multipoles ($l>29$). We limited the likelihood to the so-called lite version which contains 613 data points and marginalises over the instrumental and foreground effects.

\section{\label{sec:likelihood}Bayesian inference and likelihood analysis}

We constrained the cosmological parameters of our model, $\{\zeta,\lambda,\omega_b,\omega_c,H_0\}$, by reconstructing the posterior probability distribution (PPD) in the parameter space. The PPD  is defined as the product of the uncorrelated likelihoods specified in Sect.~\ref{sec:Data}, depending on different combinations of datasets. We sampled the parameter space with the Monte Carlo code Monte Python \citep{MP1} and the Nested Sampling algorithm \citep{Skilling2006} through the Multinest library \citep{Feroz_2009} wrapped with PyMultiNest \citep{Buchner:2014nha}. The theoretical predictions were computed with our modified version of CLASS where we also implemented the observables defined in Sect.~\ref{sec:Data}. We adopted the GetDist package \citep{Lewis:2019xzd} to analyse and plot the resulting Monte Carlo samples.

To begin with, we estimated the accuracy of the CMB distance prior by combining the QSO dataset with the Planck prior and then with full Planck (varying $\alpha$ at recombination to take into account its effects on the anisotropies). In the former, we applied a flat prior on $\zeta\in[-400,400]$ ppm. In the latter, we also included in the parameter space the absolute calibration, $A_\textrm{planck}$, as the only nuisance parameter, and the other three relevant cosmological parameters $\{n_s,\ln10^{10}A_s,\tau_\textrm{reio}\}$, where $n_s$ is the spectral index of the primordial power spectrum whose amplitude $A_s$ is normalised at a pivot scale $k_*= 0.05$ Mpc$^{-1}$, and $\tau_\textrm{reio}$ is the optical depth to reionisation.

The results of the two Bayesian analyses are reported in Table~\ref{tab:planck_prior_accuracy} where we can see that the differences in the mean values and uncertainties for the $\zeta$ posterior are insignificant. The Planck prior slightly increases the uncertainty by $0.13\sigma$ for the posterior $\lambda$. It may be counter-intuitive that full-Planck improves the constrains despite it having more free parameters than the distance prior. Nonetheless, one can also expect that the additional constraint at recombination in the former case strengthens the limits on both posteriors.
\begin{table}[t]
\caption{Comparison between the Planck distance prior and full Planck, and comparative constraining power of the ESPRESSO data point (mean and $68\%$ limits).}
 \label{tab:planck_prior_accuracy}
\centering
\begin{tabular} {l c c}
\hline\hline
Likelihoods & $\zeta$  & $\lambda$ \\ 
& (ppm) \\
\hline
QSO + full Planck & $2\pm 130$ & $0.000\pm 0.023$ \\
QSO + Planck prior & $0\pm 140$ & $0.000\pm 0.026$ \\
\hline
QSO without ESPRESSO & $2\pm150$ & $0.000\pm0.027$\\
ESPRESSO only & $0\pm170$ & $0.000\pm0.032$ \\
\hline
\end{tabular}
\tablefoot{
Both likelihoods of the bottom panel include the Planck prior.
}
\end{table}
The response of the CMB anisotropies to the variation of the fine-structure constant can therefore be neglected in the model at stake. If we take the example of the ESPRESSO data point, giving a one ppm constraint around redshift $z=1$, the variation in $\alpha$ would be about $10$ ppm at recombination time according to Eq.~\eqref{eq:alpha_param}, which is indeed insufficient to produce detectable effects. Therefore, we did not use the full Planck likelihood because it is more computationally demanding with marginal benefit. In the rest of the analysis, we systematically combined the Planck prior with the other datasets in order to account for the  correlation between cosmological parameters including $\lambda$.

\begin{figure}[t]
\resizebox{\hsize}{!}{\includegraphics{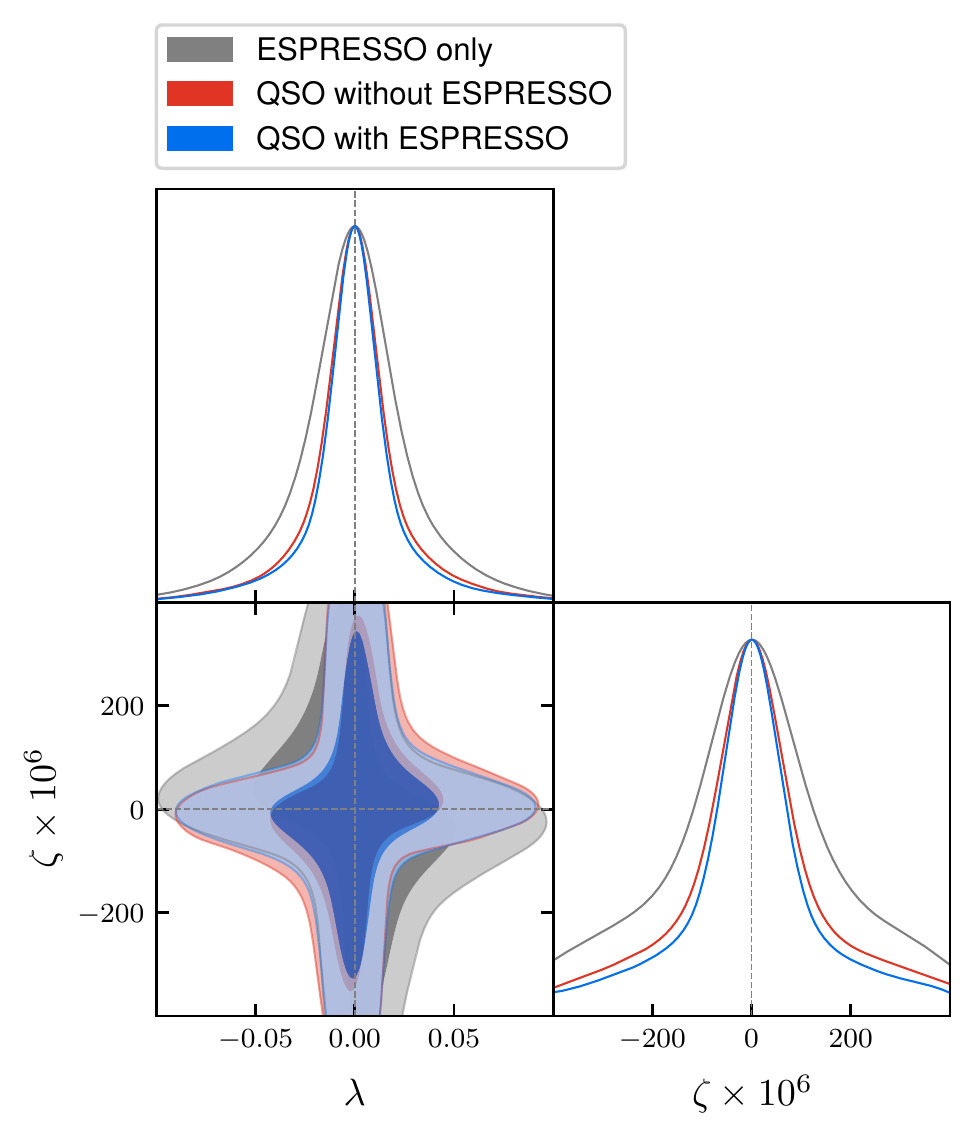}}
  \caption{Constraining power of the ESPRESSO data point with respect to QSO (+ Planck Prior in every case). Probability distribution and 2D marginalised contours (68\% and 95\% confidence level).}
\label{fig:espresso}
\end{figure}

We proceeded with a Bayesian inference that combined the QSO + Planck prior datasets with and without the recent ESPRESSO measurement in order to look at its added value in constraining the electromagnetic coupling $\zeta$ (to which we applied the same flat prior as above). The confidence contours in the $(\lambda,\zeta)$ plane in Fig.~\ref{fig:espresso} are indeed slightly improved by the addition of the this data point. We also provide the results obtained with ESPRESSO alone in the same figure. Table~\ref{tab:planck_prior_accuracy} shows that using this single measurement places constraints of the same order at $1\sigma$ confidence level.

\begin{figure}[t]
\resizebox{\hsize}{!}{\includegraphics{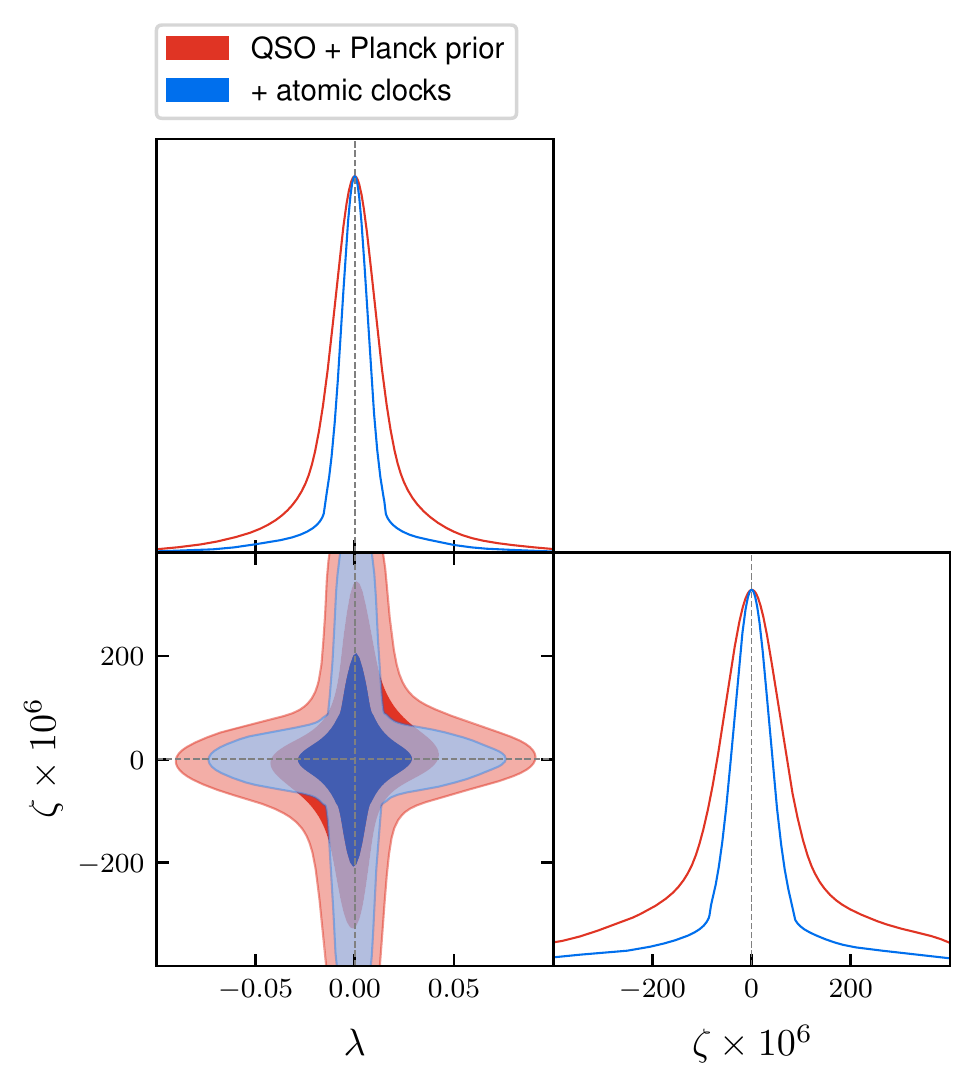}}
  \caption{Probability distribution and 2D marginalised contours (68\% and 95\% confidence level).}
\label{fig:atomic_clocks}
\end{figure}

We wanted to combine the astrophysical observations (QSO + Planck prior) with the local bounds, starting with the current $\alpha$ drift rate. The results are given in Table~\ref{tab:constraints}. The addition of the bound placed by the atomic clocks further tightens the constraints both on the electromagnetic coupling $\zeta$ and the scalar field parameter $\lambda$, as shown in Fig.~\ref{fig:atomic_clocks}. According to Eqs. \eqref{eq:drift_rate} and \eqref{eq:drift_rate_bound}, the atomic clock data constrain the product of the two parameters by approximatively
\begin{equation}
\delta(\zeta\lambda)\sim 10^{-8}/h\,,
\end{equation}
which is two orders of magnitude stronger than with astrophysical data. For example, using the ESPRESSO data point located at $z_\textrm{abs}\sim1$, Eq.~\eqref{eq:alpha_param} gives
\begin{equation}
\label{eq:espresso_constraint}
\delta(\zeta\lambda)\sim 10^{-6}/\ln2\,.
\end{equation}

\begin{figure}[t]
\resizebox{\hsize}{!}{\includegraphics{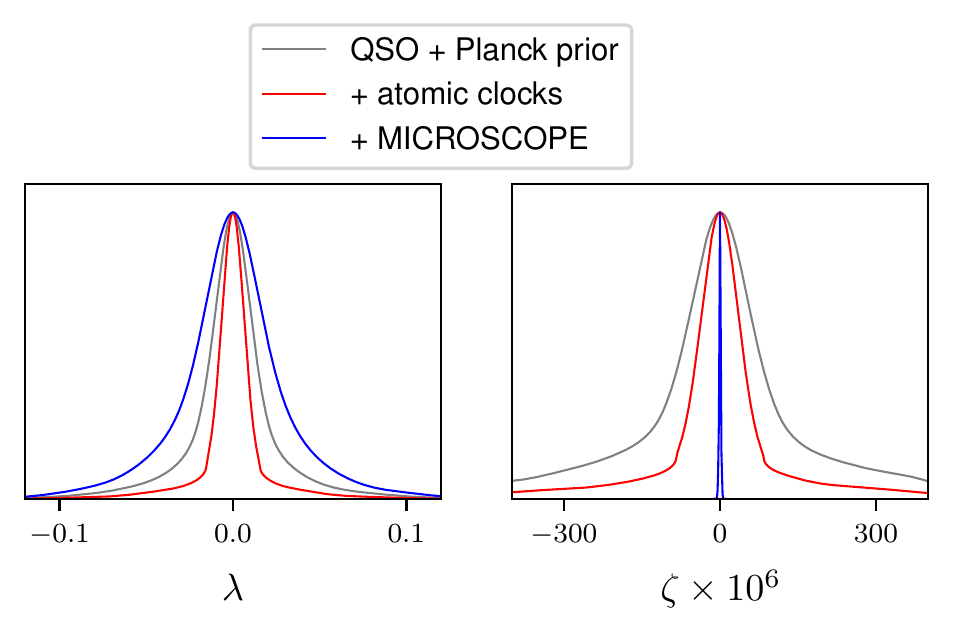}}
  \caption{Probability distribution for the two extra parameters posteriors $\lambda$ and $\zeta$.}
\label{fig:microscope}
\end{figure}

\begin{figure}[t]
\resizebox{\hsize}{!}{\includegraphics{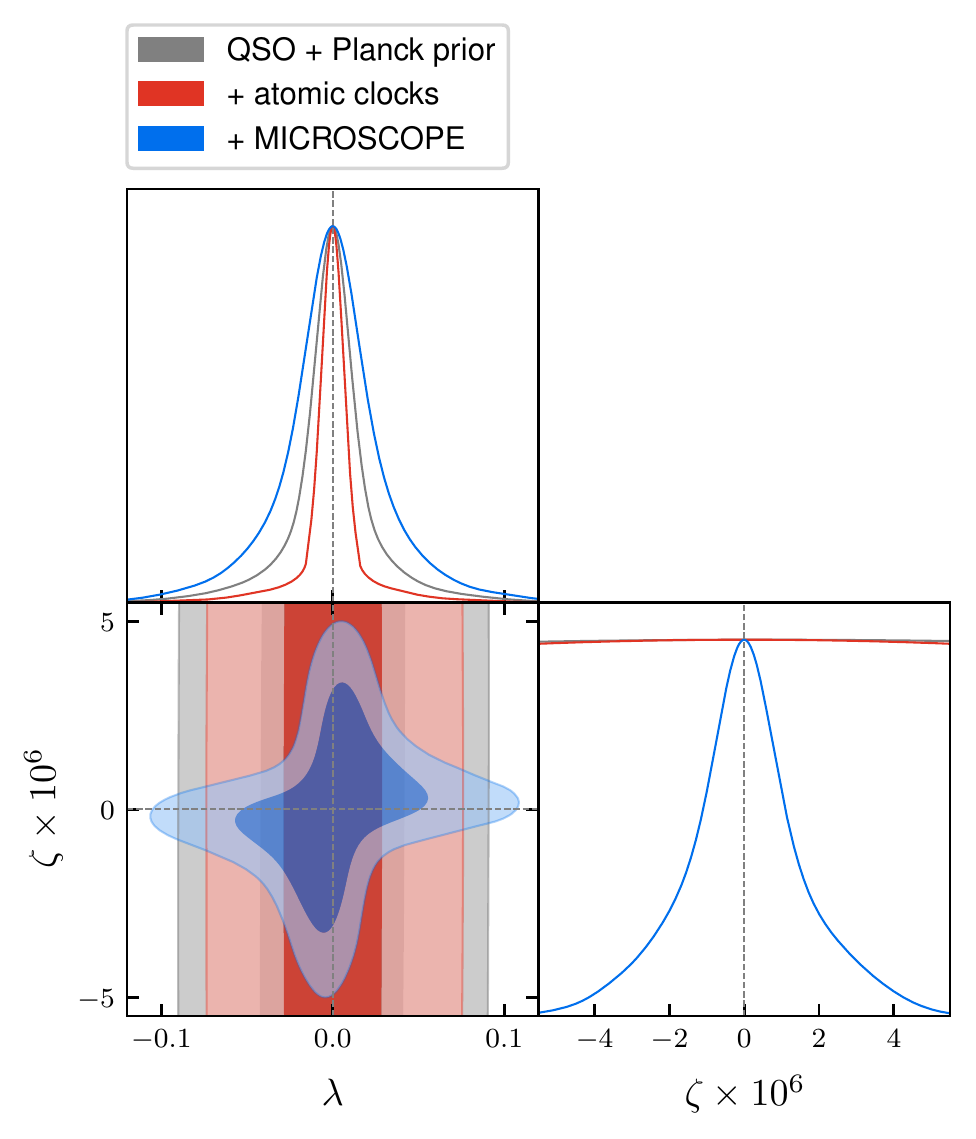}}
  \caption{Zoom onto the probability distributions and confidence contours.}
\label{fig:microscope_zoom}
\end{figure}

We supplemented the analysis with the MICROSCOPE bound, which is a Gaussian prior on $\zeta$ as mentioned earlier in Sect.~\ref{sec:Data}. Including the E\"otv\"os parameter leads to the tightest constraint on the electromagnetic coupling, $\zeta=0.0\pm 1.6$ ppm, relaxing the dark energy parameter $\lambda$ as shown in Fig.~\ref{fig:microscope}. This behaviour is similar to the findings of the forecasts in \citet{Calabrese2014} for $\Delta\alpha/\alpha$ null results. As the confidence contours are dramatically reduced for the electromagnetic coupling, as illustrated by the zoom  onto the probability distributions and confidence contours shown in Fig.~\ref{fig:microscope_zoom}, larger values of $\lambda$ are allowed in order to achieve vanishing $\Delta\alpha/\alpha$ thanks to the correlation between the two parameters encapsulated in Eq.~\eqref{eq:alpha_param}.

\begin{figure}[t]
\resizebox{\hsize}{!}{\includegraphics{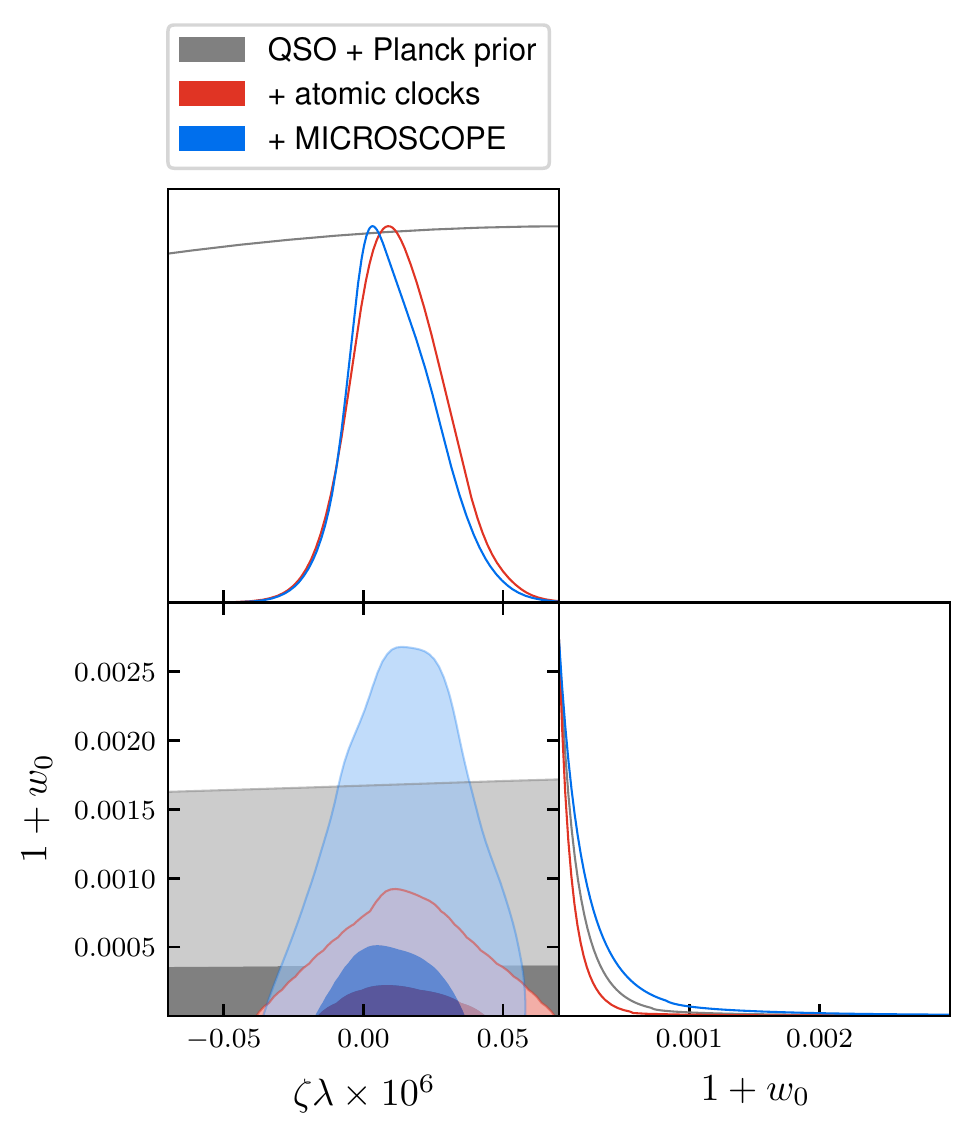}}
  \caption{$\zeta\lambda$ and dark energy equation of state posteriors. Probability distribution and 2D marginalised contours (68\% and 95\% confidence level).}
\label{fig:w0}
\end{figure}

Similar conclusions can be drawn by inspecting directly the posterior on the product $\zeta\lambda$ which is the quantity sensitive to the variation of $\alpha$. Figure~\ref{fig:w0} shows its correlation with the dark energy equation of state $w_0$ given in Eq.\eqref{eq:w0}. Both one-sigma constraints become more stringent when the atomic clocks are included on top of the QSO+Planck prior: the limits on the dark energy equation of state change from $1+w_0<1.14\times10^{-4}$ to $1+w_0<1.13\times10^{-5}$, and on the product posterior from $\zeta\lambda=0.25^{+0.56}_{-0.63}$ to $\zeta\lambda=0.013^{+0.014}_{-0.017}$. The addition of the MICROSCOPE bound does not greatly modify the posterior distribution of the product but it affects the dark energy equation of state. Without the $\eta$ bound, the product posterior is dominated by the Planck limit on $\lambda$, allowing a wider range of values for the electromagnetic coupling. Conversely, the strong MICROSCOPE constraint on $\zeta$ increases the possible values of $\lambda$, taking the dark energy equation of state gently away from a cosmological constant, $1+w_0<2.89\times10^{-4}$. The constraints placed on $w_0$ by the CMB in our parametrisation are stronger than those forecasted for the Euclid mission \citep{refId0} in \citet{Martinelli2021} with the CPL parametrisation. Consequently, we obtain weaker constraints on $\zeta$ in light of the above discussion.
\begin{table}[t]
\caption{Constraints on the two additional parameters and their product (mean and $68\%$ limits) obtained with different dataset combinations.}
 \label{tab:constraints}
\centering
\begin{tabular} {l c c c}
\hline\hline
Likelihoods & $\zeta$  & $\lambda$ & $\zeta\lambda$ \\
& (ppm) & & (ppm) \\
\hline
QSO + Planck prior & $0\pm 140$ & $0.000\pm 0.026$ & $0.25^{+0.56}_{-0.63}$\\
\rule{0pt}{3ex}
+ atomic clocks & $0\pm 110$ & $0.000\pm 0.020$ & $0.013^{+0.014}_{-0.017}$\\
\rule{0pt}{3ex}
+ MICROSCOPE & $0.0\pm 1.6 $ & $0.000\pm 0.032$ & $0.011^{+0.013}_{-0.017}$\\
\hline
\end{tabular}
\end{table}
\section{\label{sec:conlusions}Conclusions}

Current astrophysical probes of the stability of fundamental constants associated with local tests allow us to constrain dark energy models with varying $\alpha$. We constructed a simple parametrisation for this kind of model that adds just two degrees of freedom beyond $\Lambda$CDM, under the assumption that the scalar field fuelling dark energy is also driving the time variation of the fine-structure constant. As with the CPL parametrisation, the limited number of parameters is key in the ability of a given model to be efficiently constrained by observations.

We used the most recent QSO measurements in combination with CMB data. The astrophysical probes therefore combined low-redshift with high-redshift observations to obtain complementary constraints from the late and early Universe on the entire parameter space of our model. We confirm that the sensitivity of the CMB anisotropies to the electromagnetic coupling can be neglected to good accuracy, enabling us to use a simple distance prior instead of a comprehensive Planck likelihood in the Bayesian inference, and save significant computational time.

We put constraints on the parameter $\lambda$ responsible for the quintessence evolution and on the coupling $\zeta$ with the electromagnetic sector. Regarding the parametric evolution of the scalar field as a function of redshift, the constraints obtained on $\lambda$ are compatible with null variation, or in other words, with a cosmology with a cosmological constant as dark energy. As it is the product $\zeta\lambda$ that is being constrained in the parametrisation, the tight limits on $\lambda$ loosen the constraints on $\zeta$ through the degeneracy between the two parameters, because as $\lambda\rightarrow0$ any coupling value is capable of reproducing the null results of varying $\alpha$. A similar feature was found with the CPL parametrisation of the dark energy equation of state \citep{Calabrese2014}.

Despite the differences in the uncertainties, every data combination suggests a vanishing electromagnetic coupling, which corresponds to null results and is consistent with a stable fine-structure constant. While the new ESPRESSO data point improves the astrophysical limits on $\zeta$, it is the local constraints that largely dominate, particularly those from the violation of the weak equivalence principle \citep[see also][]{martins2022}. The addition of the strong E\"otv\"os parameter bound placed with the MICROSCOPE satellite gives $\zeta=0.0\pm 1.6$ ppm which is two orders of magnitude more stringent than the constraints obtained from the astrophysical probes in combination with the atomic clocks. The stronger constraint on $\zeta$ loosens those on $\lambda$ and allows slightly higher values for the dark energy equation of state. For the model considered here, the precision in the ESPRESSO measurement needs to increase by two orders of magnitude in order to be competitive with the atomic clocks. However, as the astrophysical measurements correspond to higher redshifts, this estimate is clearly model dependent. Beyond the model assumed in the present paper, ESPRESSO is essential in producing precise measurements in a redshift regime not tested by other probes. It would be interesting to evaluate how these conclusions hold when considering models involving departures from this simple parametrisation for the evolution of the field.

\section*{Acknowledgements}
\tiny

This work was financed by FEDER--Fundo Europeu de Desenvolvimento Regional funds through the COMPETE 2020--Operational Programme for Competitiveness and Internationalisation (POCI), and by Portuguese funds through FCT - Funda\c c\~ao para a Ci\^encia e a Tecnologia under projects POCI-01-0145-FEDER-028987,  PTDC/FIS-AST/28987/2017, PTDC/FIS-AST/0054/2021 and EXPL/FIS-AST/1368/2021, as well as UIDB/04434/2020 \& UIDP/04434/2020, CERN/FIS-PAR/0037/2019, PTDC/FIS-OUT/29048/2017.
MTM acknowledges the support of the Australian Research Council through Future Fellowship grant FT180100194.
The INAF authors acknowledge financial support of the Italian Ministry of Education, University, and Research with PRIN 201278X4FL and the "Progetti Premiali" funding scheme.
TMS also acknowledges the support from the Swiss National Science Foundation (SNSF) and the University of Geneva.
M-RZO acknowledges funding under project PID2019-109522GB-C51 of the Spanish Ministerio de Ciencia e Investigaci\'on.
ASM and JIGH acknowledge financial support from the Spanish Ministry of Science and Innovation (MICIN) project PID2020-117493GB-I00. ASM also acknowledges financial support from the Spanish Ministry of Science and Innovation (MICINN) under 2018 Juan de la Cierva program IJC2018-035229-I, as well as from the Government of the Canary Islands project ProID2020010129.

\bibliographystyle{aa}
\bibliography{vdf}

\end{document}